\documentclass[aps,pre,twocolumn,groupedaddress,longbibliography,showpacs]{revtex4-1}
\usepackage{amsmath,amssymb,graphicx,color,epstopdf}

\newcommand{\PP}{ \mathbf{Pr} }
\newcommand{\up}{\uparrow}

\begin{document}


\title{Birdsong dialect patterns explained using magnetic domains}


\author{James Burridge}
\email[]{james.burridge@port.ac.uk}

\author{Steven Kenney}
\affiliation{Department of Mathematics, University of Portsmouth, Portsmouth, PO1 3HF, United Kingdom}


\date{\today}

\begin{abstract}
The songs and calls of many bird species, like human speech, form distinct regional dialects. We suggest that the process of dialect formation is analogous to the physical process of  magnetic domain formation. We take the coastal breeding grounds of the Puget Sound white crowned sparrow as an example. Previous field studies suggest that birds of this species learn multiple songs early in life, and when establishing a territory for the first time, retain one of these dialects in order to match the majority of their neighbours. We introduce a simple lattice model of the process, showing that this matching behaviour can produce single dialect domains provided the death rate of adult birds is sufficiently low. We relate death rate to thermodynamic temperature in magnetic materials, and calculate the critical death rate by analogy with the Ising model. Using parameters consistent with the known behavior of these birds we show that coastal dialect domain shapes may be explained by viewing them as low temperature ``stripe states''.
\end{abstract}


\maketitle

\section{Introduction}

Under certain conditions physical systems of interacting particles form spatially ordered states. An old and famous example is the spontaneous magnetization of certain metals as they cool below the Curie temperature. The classic ``Ising model'' of this process also serves as a simple description of the appearance of order in social systems \cite{Cas09}. Whereas the directional alignment of neighbouring atomic spins within ferromagnetic materials produces an energetically favourable state, individuals in society find that aligning their behaviours with those of their neighbours can provide social benefits, or indicate the intrinsic fitness of the individual. In this paper, we introduce an ornithological application of this analogy.

Almost all birds sing or call. In some species these sounds are encoded in the DNA, while others learn \cite{Mar04}. From whom they learn and at what point in their lives varies: some birds fix their songs early in life, some continue learning throughout. Some learn from their parents and some from neighbours. Many song birds, but also some hummingbirds \cite{Yan07} and parrots \cite{Bak03}, exhibit vocal dialects: the nature of their songs or calls varies geographically with well defined and sharp boundaries between dialect domains. Because the learning mechanisms and social behaviours of different species vary considerably, so the nature and dynamics of dialect domains is also rather varied. However, the fundamental ingredients of copying, set against randomizing influences such as dispersion and death suggest that an analogy to atomic alignment and thermal noise (or random motion) may be a useful way to understand the geographical distribution of birdsong dialects. In this paper we will show that such an analogy can provide analytical insight into field observations (in particular by Nelson \cite{Nel00}) of a particular species: the Puget Sound white crowned sparrow.

\begin{figure}
\includegraphics[width=7cm]{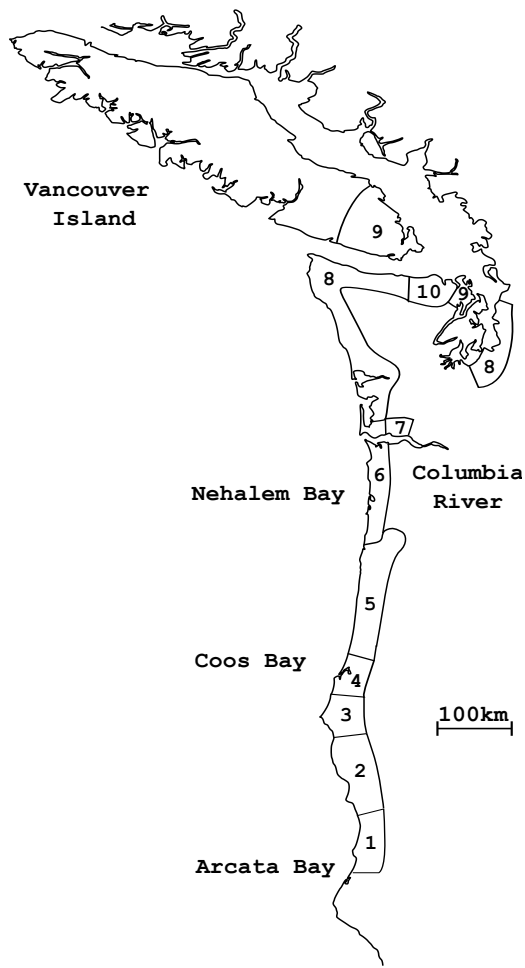}
\caption{Approximate geographical extent of song dialects of the Puget Sound white crowned sparrow along the Pacific Northwest coast of North America. Numbering used to label distinct dialects.  Figure redrawn from original data in \cite{Nel00, Nel04, Bap77,Gri44} and coastline data from the OpenStreetMap Foundation \textcircled{c} OpenStreetMap contributors. \label{dialectMap}}
\end{figure}

During the breeding season, Puget Sound white crowned sparrows occupy territories along the Pacific Northwest coast of North America, having flown 500-1900km from their wintering grounds in California \cite{Nel00}. Their songs may be analysed using audio spectograms, and the first such study (in 1977) of this species \cite{Bap77} revealed a number of distinct dialects, defined by the nature of their terminal trills, occupying well defined geographical regions along the coast. Subsequent studies have revealed the dialects illustrated on the map in Figure \ref{dialectMap}.

Male Puget sound white crowned sparrows are territorial, and adults returning to the breeding grounds typically reoccupy the territory they used the previous year and also sing the same song from year to year. Field observations and recordings suggest that male juvenile birds are able to learn more than one dialect in their hatching year, as they visit surrounding domains that are often separated by more than 100km \cite{Nel00}.  When returning to the breeding sites in the following year these yearlings ``overproduce'' \cite{Nel94}, initially singing more than one (typically two) dialects. The majority of existing adult territory holders arrive earlier than yearling males, so that when yearlings return and select an available territory, a number of neighbouring birds will be audible. Field observations and playback experiments \cite{Nel00} demonstrate that new territory holders who overproduce selectively discard dialects so as to match the majority dialect in their vicinity. Once a final selection has been made by a yearling, it sings the same song in subsequent years. It is hypothesized by ornithologists that this combination of overproduction and selective attrition gives rise to the observed dialect domains. We present a simple lattice model of this process which we use to quantify the conditions under which domains form. Using insights from the statistical physics of magnetism, we show that surface tension at domain boundaries and the formation of stripe states can be used to explain field observations of domain structure.

\section{The model}

\begin{figure}
\includegraphics[width=7.5cm]{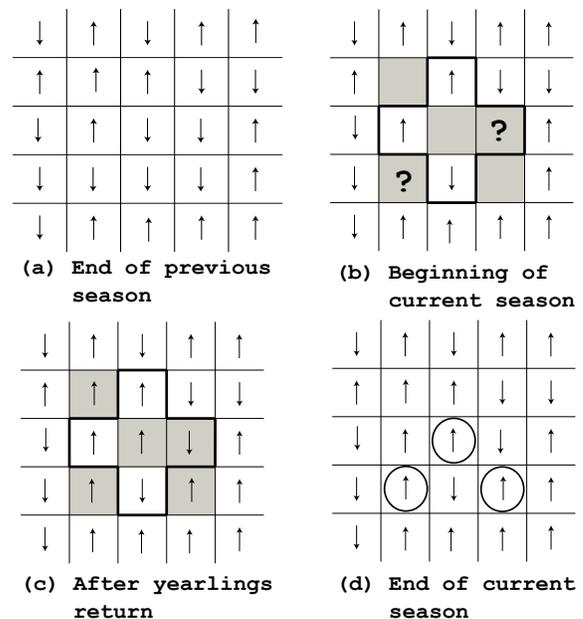}
\caption{Schematic representation of the lattice model with single phase reoccupation. Adults song states are either ``up'' $\uparrow$ or ``down'' $\downarrow$. Grey squares represent territories vacanted due to the death of an adult bird. The heavy line traces the audible neighbourhood of the central site in the four neighbour version of the model. Question marks indicate yearling birds given equivocal information about local song type. States that are changed relative to the previous season are circled. \label{lattice}}
\end{figure}

We represent territories as squares on a regular lattice, as in Figure \ref{lattice}. On the ground, the average distance between the centres of territories is $\approx 115$m \cite{Nel15}.
Since dialects extend over considerable distances, then at the boundary between two domains, typically only two different songs will be sung by the majority of adults so the competition between dialects locally involves two choices. We therefore consider two adult song types, ``up'' $\uparrow$ and ``down'' $\downarrow$, thereby simplifying the analysis of our model. At the end of the breeding season, each site in our system is occupied either by a returning adult bird, or by a yearling who has settled upon his final song type, to be sung every following year until his death. Each year, each adult bird has a probability $\alpha$ of failing to make it back to the breeding grounds after wintering in California. The average lifespan of a bird is $\approx 2.5$ years, which translates to a yearly death probability of $\alpha = 0.4$, if we approximate the lifespan as a geometric random random variable. The process of repopulating the lattice each breeding season takes place in stages. In the first stage, all surviving adults reoccupy their previous territories as shown in Figure \ref{lattice}. Non-returning adults are represented by grey squares in Figure \ref{lattice}, and we assume that the population is stable so that all of these territories will be occupied by yearlings. We consider two possible processes of reoccupation:

\subsection{Single phase reoccupation}

In the first version of the model, we assume that yearling birds listen only to preexisting adults, and ignore each other. A (possible) justification of this assumption is that yearling birds will sing more than one dialect, therefore providing no clear signal to other birds. Under this assumption, when a grey site is repopulated, the yearling in that site can only ``see'' the adults around it. In this case we assume that, based on field observations \cite{Nel00}, the yearling will match the majority of his adult neighbours. If neither dialect is in the majority then the yearling will chose between the two available dialects with equal probability.  We consider two different possibilities for the audible neighbourhood: the four nearest neighbour territories (outlined in bold for the central site in Figure \ref{lattice}), or the eight territories which form a square around each site.

\subsection{Multiple phase reoccupation}

Following the arrival of adult birds, the majority yearling birds then arrive within a window of approximately one month \cite{Nel00}. The median time to discard non-matching dialects is then 12 days, potentially allowing new territory holders to influence later arrivals. In the second version of the model, we explicitly account for the sequential arrival of yearlings. We divide the reoccupation into $n$ phases. At each phase, a fraction $1/n$ of the newly available sites are populated and each new territory holder settles on their final dialect before the next phase arrives. Subsequent phases are then able to make use of the songs of the earlier yearlings in deciding their adult song. In each case, the decision process is identical to the single phase model, except that newly arrived birds can ``see'' both the preexisting adults, and yearlings from previous phases. In the limit $n \rightarrow \infty$ we have the \emph{pure sequential} model, when yearlings arrive one by one, mimicking the physical process of cooperative random sequential adsorption \cite{Eva93}.

\section{Simulations}

\subsection{Investigation of critical death rate}

\begin{figure}
\includegraphics[width=7.5cm]{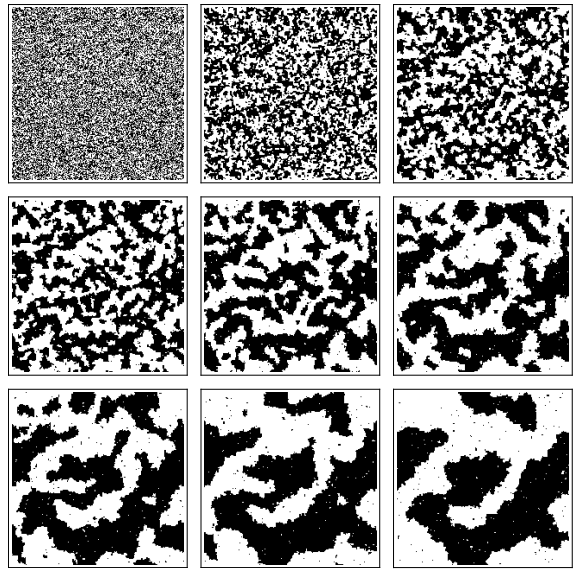}
\caption{Evolution of the single phase four neighbour model with $\alpha=0.3$ on a $200 \times 200$ lattice with periodic boundary conditions, starting from an equal proportion of dialects with no spatial order. Black and white squares represent, respectively, the dialects $\uparrow$ and $\downarrow$. System is shown after $10 \times (2^k-1)$ iterations for $k \in \{0,1,\ldots, 8\}$. \label{nonPhasedEvo}}
\end{figure}

\begin{figure}
\includegraphics[width=7.5cm]{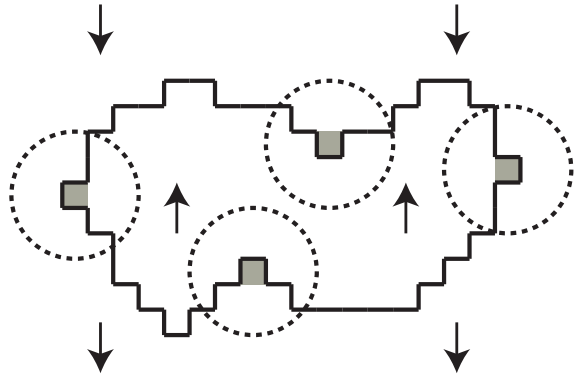}
\caption{ Example of a dialect domain. Sites that are particularly vulnerable to switching dialect after the next breeding season circled. The sites within the $\uparrow$ domain that are most likely to be lost to the $\downarrow$ domain lie on the convex parts of its boundary, whereas the sites that are likely to be gained from the $\downarrow$ domain are adjacent to the concave parts of its boundary.\label{convex}}
\end{figure}

We consider first the single phase model on a square lattice where, initially, each site has equal probability of containing an adult bird of each dialect. Provided the death rate, $\alpha$, is sufficiently low, single dialect domains form and grow (Figure \ref{nonPhasedEvo}).
As with the kinetic Ising model \cite{Red10} evolving by Glauber dynamics \cite{Glau63}, those parts of domain boundaries that are locally convex tend to shrink, whilst those that are concave tend to grow. The overall effect is one of smoothing boundaries due to an effective ``surface tension'' \cite{Red10}.  An intuitive explanation of the effect is that locally convex parts of dialect domain boundaries are more likely to contain sites that will have a majority of birds of the other dialect as neighbours during the next breeding season, as illustrated in Figure \ref{convex}.
The majority rule represents a non-linear response to local dialects. In contrast, a proportional response would be to align to a randomly chosen neighbour as in the voter model \cite{Red10, Dor01}, so that on average dialects would be chosen in proportion to their local frequency. The voter model does not lead to the formation of domains due to a lack of ``surface tension'' \cite{Dor01} and the importance of a nonlinear learning rule has recently been recognized in a non-spatial model \cite{Pla14} of bird dialects.

By increasing the death rate of adult birds, dialect continuity from year to year is reduced: a site which at the end of one year was surrounded by a majority of one dialect will be more likely, if the death rate is high, to be converted by random deaths into a site surrounded by a majority of the other, or one where neither dialect is in the majority. As a consequence there exists a critical death rate, analogous to the Curie temperature in ferromagnets, above which domains dominated by a single dialect cannot form. To discover the critical death rate we introduce a correlation length, $\xi$, for a square lattice of side $L$ with periodic boundary conditions. Letting $S(x,y) \in \{1,-1\}$ (where $\uparrow \equiv 1$ and $\downarrow \equiv -1$) denote the dialect at site $(x,y)$ we define
\begin{equation}
\xi = \mathbb{E} \left[ \sum_{x'=1}^{L/2} S(x,y)S(x',y)\right]
\end{equation}
where $\mathbb{E}[\cdot]$ denotes the expectation over the equilibrium probability distribution of system states. Due to translational invariance $\xi$ is independent of $(x,y)$. From its definition we see that if the system consists of one single large domain then $\xi = L/2$ and if there are no spatial correlations between dialects at different sites then $\xi=0$. If the equilibrium domain size is finite and less than the system size then $\xi$ will lie between these two values.

\begin{figure}
\includegraphics[width=8cm]{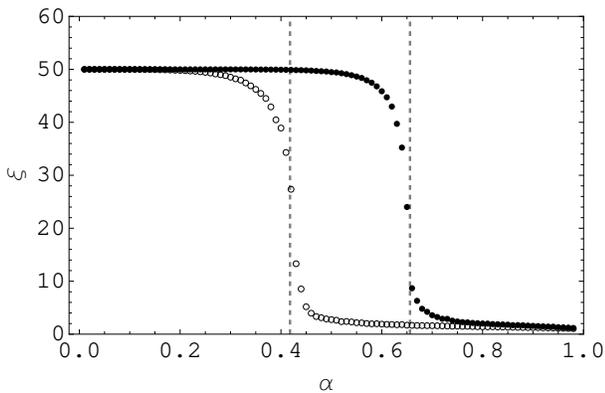}
\caption{ Correlation length as a function of death rate $\alpha$ in a $100 \times 100$ system with periodic boundary conditions. Open circles correspond to the four neighbour model, and black dots to the eight neighbour model. Dashed lines show analytical estimates of the critical death rates $\alpha_c \approx 0.4178$ (four neighbour model) and $\alpha_c = 0.657$ (eight neighbour model). \label{nonPhased}}
\end{figure}

We have estimated $\xi$ by initially setting each site to state $\uparrow$  with probability $0.95$, and setting $\alpha=0.01$ (in order to avoid stripe states \cite{Red09}). The system was then allowed to reach equilibrium and the correlation length estimated by computing the sum
\begin{equation}
\sum_{x'=1}^{L/2} S(x,y)S(x',y)
\end{equation}
at every site in the system at a sequence of times, before computing a spatial and time average. Following this the death rate was incrementally increased, and the system allowed to equilibrate before re-calculating $\xi$. The results of this procedure are shown in Figure \ref{nonPhased} for both the four and eight neighbour versions of the model.
From the Figure we see that for sufficiently low death rate the system consists of a single dialect domain. However, once $\alpha$ reaches a critical level the correlation length drops rapidly indicating fragmentation into a disordered state. The critical death rate is considerably higher in the eight neighbour model, because a greater fraction of neighbours need to die in order to change the majority dialect around a cell. However, field observations \cite{Nel00} show that the number of neighbours available to learn from varies between one and four, suggesting that the four neighbour model more plausibly characterises the real system. Of particular interest is the fact that the critical death rate in the four neighbour model is close to the actual death rate of the Puget Sound white crowned sparrow. Such a death rate makes dialects only marginally stable in this model, suggesting that the extra information provided on local dialects to later arriving yearlings by their early arriving counterparts may play an important role. To quantify this effect we simulated the multiple phase four neighbour model in the cases $n=2$ and $n \rightarrow \infty$, in order to determine the correlation length as a function of death rate.
\begin{figure}
\includegraphics[width=8cm]{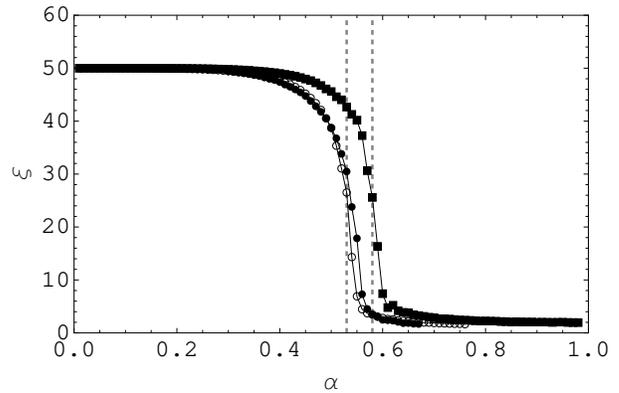}
\caption{Correlation length as a function of death rate $\alpha$ in a $100 \times 100$ system with periodic boundary conditions. Open circles correspond to the four neighbour two-phase model and black dots to the corresponding Ising model. Squares correspond to the fully sequential model. Dashed lines show estimated critical values $\alpha_c \approx 0.53$ (two phase) and $\alpha_x \approx 0.58$ (fully sequential).  \label{phasedXi}}
\end{figure}
The results are presented in Figure \ref{phasedXi}. We see that in both cases the critical death rate is considerably higher than in the single phase model, so that the field observed death rate $\alpha \approx 0.4$ would lead to stable dialects.

\subsection{Stripe States}

Provided that $\alpha < \alpha_c$ then (as with the Ising model when $T<T_c$) a system consisting of a single dialect is stable against the invasion of the other song. To see this, suppose that the dominant dialect is $\uparrow$ and consider the fate of a closed ``droplet'' domain of $\downarrow$ within the system. Due to surface tension, the droplet will evolve toward a circular shape, which will then shrink due to its convexity. Despite the fact that the single dialect state is absorbing in the sense that the system cannot escape it, it has been shown in the case of the Ising model, that the system can freeze into stable states containing multiple domains of opposing states \cite{Red01,Red09}. These stable domains consist of ``stripes'' which, because of their straight boundaries, lack surface tension which could shrink or expand them. The probability that such states will form in a rectangular system, starting from a randomized initial condition, depends strongly on its aspect ratio. Higher aspect ratios tends to produce multiple stripes, stretching across the short side of the system.
\begin{figure}
\includegraphics[width=8.5cm]{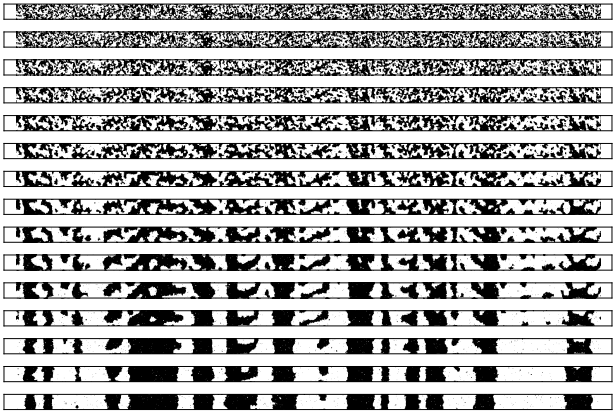}
\caption{Evolution of the sequential four neighbour model with $\alpha=0.4$ on a $2000 \times 50$ lattice with wall boundary conditions at long (horizontal sides) and periodic boundary conditions at either end. Black (white) squares represent dialects $\uparrow$ ($\downarrow$). System is shown after $10 \times \lfloor 1.5^k-1 \rfloor$ iterations for $k \in \{1,2,\ldots, 15\}$. \label{sequentialStrip}}
\end{figure}
The existence of such ``frozen'' dialect states would provide a plausible explanation for the structure of the Puget Sound white crowned sparrow's dialect domains, because their coastal breeding grounds form a long strip. To investigate the typical sizes of such domains in our dialect model we have simulated the fully sequential version on a high aspect ratio rectangle (Figure \ref{sequentialStrip}). For large systems, when domains become large, the dynamics of their boundaries may be approximated with continuous curves and the explicit length scale of individual sites becomes irrelevant \cite{Red10}. In the absence of an absolute length scale only the relative sizes of structures within the system are meaningful. We may exploit this scale free property of large systems to compare the aspect ratio of stripe domains in our simulation to those of real dialect domains. By coarse graining the system of Figure \ref{sequentialStrip} into $10 \times 10$ cells, we estimated the aspect ratios $AR$ of single domains to lie in the range $AR \in [0.6,3.5]$ with mean and standard deviation $\mu(AR)=1.4$ and $\sigma(AR) = 1.1$. Distribution maps \cite{Gri44} suggest that breeding territories of Puget Sound white crowned sparrow spread $\approx 50$km inland and combined with the dialect distributions mapped in Figure \ref{dialectMap}, we predict that the aspect ratios of the real dialect domains lie in the range $AR \in [1,4.3]$ with mean and standard deviation $\mu(AR)=2.8$ and $\sigma(AR) = 1.6$. The similarity between the simulated and observed distributions supports our hypothesis that the observed domains arise due to the processes modeled, and are stripe states.

\section{Approximation with Kinetic Ising Model}

Our analysis has made use of an analogy between dialect domains in bird populations and domains of aligned spins in ferromagnets. We now make this connection explicit in the single phase case by deriving Ising models with transition rates which match the expected transition rates in our dialect system. This enables us to derive an approximate relationship between death rate and thermodynamic temperature. We also demonstrate how phased arrival of yearling birds may be viewed as extending the distance over which nearby adult dialects can be perceived.

\subsection{Single Phase Reoccupation}

\begin{figure}
\includegraphics[width=8.5cm]{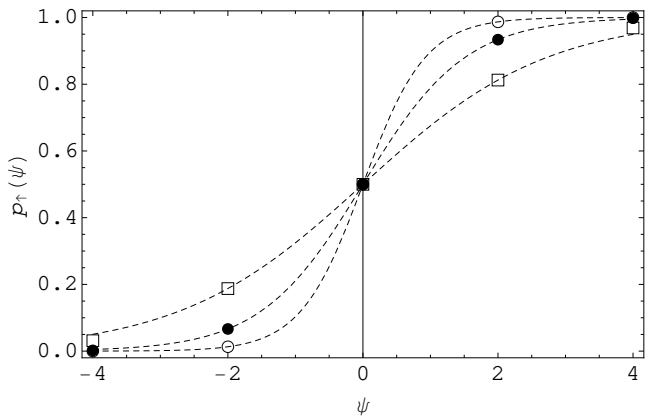}
\caption{The discrete function $p_{\uparrow}(\psi)$ for $\alpha \in \{0.1,0.3,0.6\}$ (open circles, filled circles, squares). Also shown as dashed lines are the continuous functions $\omega_{\uparrow}(\psi) = (1+\tanh[\beta(\alpha) \psi])/2$. \label{pPsi}}
\end{figure}

Consider the four neighbour model. Letting $\langle x,y \rangle$ denote the nearest neighbours of site $(x,y)$ we define
\begin{equation}
\psi(x,y) = \sum_{\langle x,y \rangle} S(x',y'),
\end{equation}
where $S(x',y') \in \{1,-1\}$ is the dialect at site $(x',y')$ at the end of a breeding season. We now suppose that at the start of the following season, site $(x,y)$ is vacant. The yearling male occupying this site will not typically have access to $\psi(x,y)$ due to the non-return of one or more of its neighbours. This disruption of information plays the same role as temperature in the Ising model. We now define the indicator function for the return of an adult bird at each site
\begin{equation}
R(x,y) = \begin{cases}
1 &\text{ with prob. } 1-\alpha \\
0 &\text{ with prob. } \alpha.
\end{cases}
\end{equation}
Using this indicator, we can compute the sum of the dialects seen by the new territory holder at $(x,y)$
\begin{equation}
\tilde{\psi}(x,y) = \sum_{\langle x,y \rangle} R(x',y') S(x',y').
\end{equation}
The yearling will chose his dialect according to the sign of this ``noisy'' version of $\psi(x,y)$. We now define the step function
\begin{equation}
f(z) = \begin{cases}
1 & \text{ if } x>0 \\
\frac{1}{2} & \text{ if }x=0 \\
0& \text{ if }  x<0.
\end{cases}
\end{equation}
Given $\psi(x,y)$, the probability that he will choose the $\uparrow$ dialect is
\begin{align}
p_{\uparrow}(\psi) &= \PP\{\tilde{\psi}>0\} + \tfrac{1}{2}  \PP\{\tilde{\psi}=0\}\\
&= \mathbb{E}[f(\tilde{\psi})]
\label{pup}
\end{align}
where the expectation is taken over all possible combinations of vacated sites. For notational simplicity we have omitted the arguments $(x,y)$ of $\psi$ and $\tilde{\psi}$. To compute this expectation as a function of $\alpha$ we note that the numbers $N_{\uparrow}, N_{\downarrow}$ of birds with dialects $\uparrow$ and $\downarrow$ around site $(x,y)$ at the \emph{start} of the breeding season (allowing for vacated territories), conditional on $\psi$, are binomially distributed
\begin{multline}
\PP\{N_{\uparrow} = u \cap N_{\downarrow}=v | \psi\}  \\
= \binom{2 +\tfrac{\psi}{2}}{u}\binom{2 -\tfrac{\psi}{2}}{v}\alpha^{4-u-v} (1-\alpha)^{u+v}.
\end{multline}
Noting that $\tilde{\psi} = N_{\uparrow} - N_{\downarrow}$ we can compute the expectation in (\ref{pup}) as
\begin{equation}
\mathbb{E}[f(\tilde{\psi})] = \sum_{u=0}^{2+\frac{\psi}{2}} \sum_{v=0}^{2-\frac{\psi}{2}}\PP\{N_{\uparrow} = u \cap N_{\downarrow}=v | \psi\} f(u-v),
\end{equation}
leading to the following expression for $p_{\uparrow}(\psi)$
\begin{equation}
p_{\uparrow}(\psi)=
\begin{cases}
\frac{\alpha^4}{2} & \text{ if } \psi=-4 \\
\frac{ 3 \alpha^2}{2} - 2 \alpha^3 + \alpha ^4 & \text{ if } \psi=-2 \\
\frac{1}{2} & \text{ if } \psi=0 \\
 1 - \frac{ 3 \alpha^2}{2} + 2 \alpha^3 - \alpha ^4 & \text{ if } \psi=2 \\
1 - \frac{\alpha^4}{2} & \text{ if } \psi=4
\end{cases}
\end{equation}
As shown in Figure \ref{pPsi} $p_{\uparrow}(\psi)$ takes the form of a smoothed discrete version of $f(\psi)$.
Now let us interpret dialect states at the end of each season as Ising spins, evolving under Glauber dynamics \cite{Glau63} at inverse thermodynamic temperature $\beta$. In this case sites are selected one at a time uniformly at random from the lattice, and the probability that a selected spin will be set to the $\uparrow$ state is
\begin{equation}
\omega_{\uparrow}(\psi) := \frac{1 + \tanh(\beta \psi)}{2},
\end{equation}
which is also a smoothed discrete version of $f(\psi)$. If we now choose $\beta$ so that $p_{\uparrow}(\psi)$ and $\omega_{\uparrow}(\psi)$ match for $\psi \in \{-2,0,2\}$, then any site on a domain boundary will have a probability of changing state the next time it is updated which is identical between the two models, given the value of $\psi$. It is important to realize however that the simultaneous updates that take place in the dialect model introduce correlations between state flips which are not present in Glauber dynamics. Matching the values of $p_{\uparrow}(\psi)$ and $\omega_{\uparrow}(\psi)$, we find that
\begin{equation}
\beta(\alpha) = \frac{1}{2} \tanh^{-1}(1-3\alpha^2+4 \alpha^3-2 \alpha^4).
\end{equation}
We may now make use of the approximate analytical relationship between death rate and inverse temperature to estimate the critical death rate in the dialect model. Equating $\beta(\alpha)^{-1}$ to the exact critical temperature of the Ising model \cite{Kra41}
\begin{equation}
\frac{1}{\beta(\alpha)} = \frac{2}{\ln(1+\sqrt{2})}
\end{equation}
and solving for $\alpha$, gives
\begin{equation}
\alpha_c \approx 0.4178
\end{equation}
to four significant figures. Simulations of the correlation length $\xi$ demonstrate that this method of prediction is remarkably accurate in the four neighbour case (Figure \ref{nonPhased}).

For the eight neighbour model, we redefine $\psi(x,y)$ to be the sum of the eight dialects relevant to site $(x,y)$. In this case
\begin{multline}
\PP\{N_{\uparrow} = u \cap N_{\downarrow}=v | \psi\}  \\
= \binom{4 +\tfrac{\psi}{2}}{u}\binom{4 -\tfrac{\psi}{2}}{v}\alpha^{8-u-v} (1-\alpha)^{u+v}.
\end{multline}
We let $p_{8\uparrow}(\psi)$ represent the probability of a new territory holder choosing the $\uparrow$ dialect in the eight neighbour model. Repeating the calculation steps above we find that
\begin{align}
p_{8\uparrow}(0) &=
\frac{1}{2} \\
p_{8\uparrow}(2) &= 1- 5 \alpha^2 + 20 \alpha^3 - 45 \alpha^4 - \frac{115 \alpha^6}{2} + 30 \alpha^7 - 7 \alpha^8 \\
p_{8\uparrow}(4) &= 1- \frac{15 \alpha^4}{2} + 24 \alpha^5 - 34 \alpha^6 + 24 \alpha^7 - 7 \alpha^8 \\
p_{8\uparrow}(6) &= 1- \frac{7 \alpha^6}{2} + 6 \alpha^7 - 3 \alpha^8 \\
p_{8\uparrow}(8) &= 1- \frac{\alpha^8}{2}
\end{align}
with $p_{8 \uparrow}(\psi) = 1- p_{ 8\uparrow}(-\psi)$. Matching $p_{8\uparrow}(\psi)$ and $\omega_{\uparrow}(\psi)$ for $\psi \in \{-2,0,2\}$ we have
\begin{equation}
\beta_8(\alpha) = \frac{1}{2} \tanh ^{-1}(2 p_{8\uparrow}(2) -1 ).
\end{equation}
Equating this quantity with the series estimate \cite{Dom66,Fan69} $T_c \approx 5.2599$ of the critical temperature in the eight neighbour Ising model we obtain
\begin{equation}
\alpha_{c} \approx 0.657.
\end{equation}
This may be compared to our simulation estimate $\alpha_{c} \approx 0.651$ (inflection point of $(\alpha,\xi)$ graph in Figure \ref{nonPhased}). Whilst still an effective estimation method, the slight reduction in accuracy may, we suggest, be related to that fact that not all sites on domain boundaries have identical probabilities of changing state between the two models.

\subsection{Multiple Phase Reoccupation}

\begin{figure}
\includegraphics[width=8.0cm]{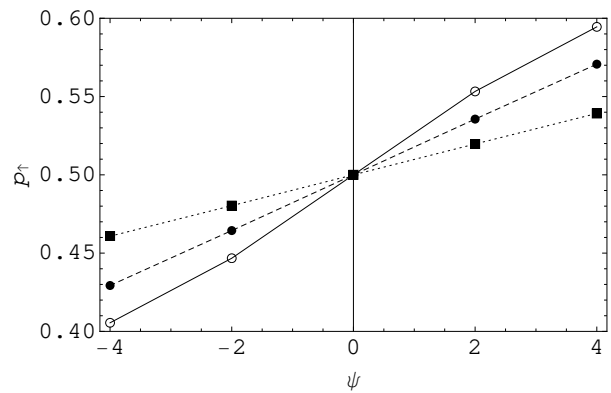}
\caption{The influence of second and third ring sites in the case where $\psi_1=0$ and $\alpha=0.4$ (field observed death rate). Open circles: $p_{\up}(0,\psi,\psi)$, closed circles: $p_{\up}(0,\psi,0)$, squares: $p_{\up}(0,0,\psi)$  \label{twoPhaseProbs}}
\end{figure}

We now show how the phased arrival of new territory holders may be seen to induce an extended-range interaction. We consider the two phase, four neighbour model.
\begin{figure}
\includegraphics[width=5.0cm]{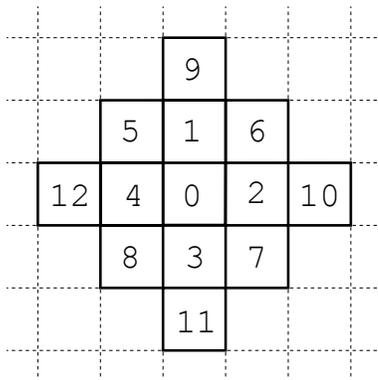}
\caption{Sites, outlined in bold, that can influence the dialect of the central site (0) in the two phase model.  \label{ELR}}
\end{figure}
In this case a new territory holder may be influenced by the states of the set of sites which lie within two lattice steps of his territory. With reference to Figure \ref{ELR} we let $\psi_1, \psi_2$ and $\psi_3$ be the sums of dialects in, respectively, the set of nearest, next nearest, and next next nearest neighbour sites of a new territory holder
\begin{align}
\psi_1 &:= S_1 + S_2 + S_3 + S_4 \\
\psi_1 &:= S_5 + S_6 + S_7 + S_8 \\
\psi_1 &:= S_9 + S_{10} + S_{11} + S_{12}.
\end{align}
We refer to these sets of sites as the first, second and third ``rings''. If a new territory holder at site 0 is part of the first phase then the probability that he will select the $\uparrow$ dialect is identical, given $\psi_1$, to the four neighbour single phase model. However, if he is part of the second phase, then it is possible that currently vacant sites in the first ring may be occupied in the first phase. The fate of these first ring sites is decided by the states of sites in the second and third rings. Sites in even higher rings only become relevant in higher phase models. We define $p_{\uparrow}(\psi_1,\psi_2,\psi_3)$ to be the probability that a new territory holder with first, second and third ring dialect sums $\psi_1$, $\psi_2$ and $\psi_3$ will choose the $\uparrow$ dialect. This probability can be computed exactly (see appendix) if we assume that all dialect arrangements within each ring, consistent with $\{\psi_i\}_{i=1}^3$ are equally probable. To quantify the importance of the second and third ring sites we have illustrated (Figure \ref{twoPhaseProbs}) the dependence of $p_\up$ on $\psi_2$ and $\psi_3$  when the first ring sites give equivocal information $\psi_1=0$. We see that the states of these outer--ring sites can change $p_\up$ by, at most, a factor of $\tfrac{3}{2}$ when realistic death rates are assumed.

By selecting dialects uniformly at random from the lattice and updating them using the probabilities $p_{\uparrow}(\psi_1,\psi_2,\psi_3)$, which may be viewed as a form of Glauber dynamics with appropriately chosen (by rather complicated) Hamiltonian, we obtain the correlation length estimates plotted in Figure \ref{phasedXi}. From this we see that the extended range interactions induced by the phased model, although weak, create a significant increase in the stability of dialect domains against death of adults.

\section{Conclusion}

 Using field data on the behaviour and songs of the Puget Sound white crowned sparrow \cite{Nel00}, we have developed a simple lattice model which may be used to explain the large dialect domains which appear along its coastal breeding grounds. We have shown that the destabilising effect of adult death, and the song matching behaviour of yearling males may be viewed as analogous, respectively, to thermodynamic temperature and spin-spin interactions in two dimensional magnetic materials. Pursuing this analogy we have shown how stable dialect domains may be viewed as stripe states with similar size distribution to observed domains, we have calculated the maximum death rate for which dialects will persist ($ \approx 60\%$), and we have shown how phased arrival of new territory holders can significantly increase dialect stability through an effective interaction beyond nearest neighbour territories. Birdsong dialects are widely observed and take a variety of forms \cite{Pod07}; we suggest that the analogy to ordering in physical systems may be usefully applied to other species and potentially to the study of observed and historical human dialect domains.

\appendix*

\section{Calculation of two phase probabilities}

We show how to calculate the conditional probability that a new territory holder in the two phase model will choose the $\uparrow$ dialect given the first, second and third ring dialect sums $\psi_1$, $\psi_2$ and $\psi_3$. We write this probability  $p_{\uparrow}(\psi_1,\psi_2,\psi_3) \equiv \PP(\up | \psi_1, \psi_2, \psi_3)$.

We adopt the site numbering in Figure \ref{ELR}. Let $\vec{S} = (S_1, S_2, \ldots S_{12})$ represent the song states at the end of the previous breeding season, and let $\vec{R} = (R_1,R_2, \ldots R_{12})$ be the indicators for returning adults at the start of the current season and $\vec{r} = (r_1,r_2,r_3,r_4)$ be indicators of arriving yearlings in the first wave to the nearest neighbour sites of the origin. Let $\vec{B} =(B_1,B_2,B_3,B_4)$ be a vector of Bernoulli variables $B_i \in \{-1,1\}$ which indicate song decisions in these sites in the case that returning yearlings are presented with equivocal information. We have
\begin{align}
\PP(R_i=0 \cap r_i = 0 ) &= \frac{\alpha}{2} \\
\PP(R_i=0 \cap r_i = 1) &= \frac{\alpha}{2} \\
\PP(R_i=1 \cap r_i = 0) &= 1 - \alpha \\
\PP(R_i=1 \cap r_i = 1) &= 0 \\
\PP(R_i=1) &= 1-\alpha \\
\PP(R_i=0) &= \alpha .
\end{align}
We also define
\begin{align}
\psi_1(\vec{S}) &:= \sum_{k=1}^{4} S_k \\
\psi_2(\vec{S}) &:= \sum_{k=5}^{9} S_k\\
\psi_3(\vec{S}) &:= \sum_{k=9}^{12} S_k.
\end{align}
To compute the probability $p_{\uparrow}(\psi_1,\psi_2,\psi_3)$ we condition on the central site being empty and consider the two possible stages at which a yearling bird can reoccupy it.

\subsection*{Case I: Central site reoccupied in the first phase}
We define $p_1(\uparrow | \vec{S}, \vec{R})$ to be the probability (conditional on $\vec{S}, \vec{R}$) that the central site is in the $\up$ state at the end of the season given that the yearling at the site returns in the first phase. We define
\begin{equation}
X_0 := \sum_{k=1}^4 S_k R_k
\end{equation}
then
\begin{equation}
p_1(\uparrow | \vec{S}, \vec{R}) = I_{\{X_0>0\}} + \frac{1}{2} I_{\{X_0=0\}},
\end{equation}
where the indicator function $I_A$ of the event $A$ is defined
\begin{equation}
I_A = \begin{cases}
1 & \text{ if A occurs} \\
0 & \text{ otherwise }.
\end{cases}
\end{equation}

\subsection*{Case II: Central site reoccupied in the second phase}

We define $p_2(\uparrow| \vec{S}, \vec{R}, \vec{r}, \vec{B})$ to be the probability (conditional on $\vec{S}, \vec{R},\vec{r}, \vec{B}$) that the central site is in the $\up$ state at the end of the season given that the yearling at the site returns in the second phase. We define
\begin{equation}
X_k := \sum_{i \in \langle k \rangle} S_i R_i \ \ k \in \{1,2,3,4\}
\end{equation}
where $\langle k \rangle$ denotes the nearest neighbours of site $k$. The random variable $X_k$ is the sum of all states around site $k$ at the \emph{start} of the season. The quantity that determines the fate of the central site is the sum of the states in sites $\{1,2,3,4\}$ at the \emph{end} of the first phase. Let $\{s_k\}_{k=1}^4$ (note change of case) be these states where $s_k \in \{-1,0,1\}$. If $R_k=1$ then site $k$ is occupied by last year's adult and $s_k = S_k$. If $r_k=1$ then the returning bird at site $k$ is a yearling and bases its decision on $X_k$. If $R_k=r_k=0$ then the site remains empty so $s_k=0$. Therefore
\begin{equation}
s_k = R_k S_k + r_k \left[ I_{\{X_k>0\}} - I_{\{X_k<0\}} + B_i I_{\{X_k=0\}} \right].
\end{equation}
We now define
\begin{equation}
x_0 := \sum_{i=1}^4 s_i,
\end{equation}
then
\begin{equation}
p_2(\uparrow| \vec{S}, \vec{R}, \vec{r}, \vec{B}) = I_{\{x_0>0\}} + \frac{1}{2} I_{\{x_0=0\}}.
\end{equation}

\subsection*{Unconditional probability}

Since the central site is equally likely to be filled in the first or second phase, then conditional on $\vec{S}, \vec{R}, \vec{r}, \vec{B}$ we have
\begin{equation}
\PP(\uparrow| \vec{S}, \vec{R}, \vec{r}, \vec{B}) = \frac{1}{2} \left[ p_1(\uparrow | \vec{S}, \vec{R}) + p_2(\uparrow| \vec{S}, \vec{R}, \vec{r}, \vec{B}) \right].
\end{equation}
To compute the unconditional probability we define
\begin{align}
f_{Rr}(u, v) &= \PP(R_i=u \cap r_i=v ) \\
f_{R}(u) &= \PP(R_i=u )
\end{align}
The joint probability mass function of $\vec{R}$ and $\vec{r}$ is then
\begin{align}
f(\vec{u}, \vec{v}) &:= \PP(\vec{R} = \vec{u} \cap \vec{r} = \vec{v} ) \\
&= \prod_{k=1}^{4} f_{Rr}(u_k,v_k) \prod_{k=5}^{12} f_{R}(u_k).
\end{align}
We now assume that conditional on the values of $\psi_1,\psi_2,\psi_3$ all values of $\vec{S}$ are equally likely. We let $\mathbf{S}$ be the set of all possible values of $\vec{S}$ and define
\begin{multline}
A(y_1,y_2,y_3) = \\
  \{ \vec{S} \in \mathbf{S} |\psi_1(\vec{S})=y_1,\psi_2(\vec{S})=y_2,\psi_3(\vec{S})=y_3 \}.
\end{multline}
Since all 16 values of $\vec{B}$ are equally probable then
\begin{equation}
p_\uparrow(y_1,y_2,y_3) = \frac{1}{16 |A|}\sum_{\vec{S} \in A, \vec{u}, \vec{v}, \vec{B}} f(\vec{u}, \vec{v})p(\uparrow| \vec{S}, \vec{u}, \vec{v},\vec{B})
\end{equation}
where we have suppressed the arguments of $A$ for notational compactness. To compute all such probabilities requires us to sum over $2^{32}$ combinations of song states $\vec{S}$, arrival times $(\vec{R}, \vec{r})$ and decision variables $\vec{B}$. Using a simple Python program running in parallel on five cores of an eight core workstation this can be achieved in approximately one day. An example of the result of this calculation is
\begin{multline}
p_\up(0,0,2) =  -\frac{7 \alpha ^{11}}{512}+\frac{17 \alpha ^{10}}{768}-\frac{23 \alpha ^9}{768}+\frac{23 \alpha ^8}{384}-\frac{149 \alpha ^7}{1536}
\\ +\frac{17 \alpha ^6}{128}- \frac{119 \alpha ^5}{768}+\frac{23 \alpha ^4}{384}+\frac{\alpha ^3}{24}-\frac{5 \alpha ^2}{48}+\frac{\alpha }{12}+\frac{1}{2}
\end{multline}
We have verified our analytical results using Monte Carlo simulations. We note also that further verification is provided by considering the cases $\alpha=0$ and $\alpha=1$. For example
\begin{equation}
p_\up(0,0,2)\vert_{\alpha=0} = p_\up(0,0,2)\vert_{\alpha=1} = \frac{1}{2}
\end{equation}
consistent with the fact that in the limit $\alpha \rightarrow 0$, every yearling has four adult neighbours and so cannot be influenced by the outer rings, and when $\alpha=1$ all information from the previous season is lost. For $0 < \alpha < 1$, $p_\up(0,0,2)$ has a single maximum at $\alpha = 0.5677$.

\begin{acknowledgments}
The authors would like to thank Doug Nelson for providing data and further information about the Puget Sound white crowned sparrow.
\end{acknowledgments}

\bibliographystyle{unsrt}
\bibliography{}

\end{document}